\def\year{2017}\relax
\pdfoutput=1
\documentclass[letterpaper]{article} 
\usepackage{aaai17}  
\usepackage{times}  
\usepackage{helvet}  
\usepackage{courier}  
\usepackage{url}  
\usepackage{graphicx}  
\usepackage{booktabs}
\usepackage{amsmath}
\usepackage{amssymb}
\usepackage{xcolor}

\newcounter{NumberOfComments}
\stepcounter{NumberOfComments}

\newcounter{JSNumberOfComments}
\stepcounter{JSNumberOfComments}

\begin{document}
%
\title{What Gets Media Attention and How Media  Attention Evolves Over Time - \\ Large-scale Empirical Evidence from 196 Countries}
\author{Jisun An and Haewoon Kwak\\
Qatar Computing Research Institute\\
Hamad Bin Khalifa University\\
Doha, Qatar\\
\{jisun.an, haewoon\}@acm.org
}
\maketitle

\maketitle
\begin{abstract}
 It is known that news topics,  covered more frequently and over longer periods of time, are considered to be important to the public. Hence, what gets media attention and how media attention evolves over time has been studied for decades in communication study. However, previous studies are confined to a few countries or a few topics, mainly due to lack of longitudinal global data. In this work, we use a large-scale news data compiled from 196 countries to provide empirical analyses of media attention dynamics.
\end{abstract}

\section{Introduction}

Of the millions of events that  occur daily in the world, only a tiny fraction is covered by news media. 
This ``selected set of news'' represents and shapes our understanding of the world~\cite{ndlela2005african}. 
The news selection process is often guided by an understanding of news values; a set of criteria to determine newsworthiness of an event.  
Galtung and Ruge (\citeyear{galtung1965structure}) firstly presented twelve factors, including frequency, unexpectedness, and reference to the elite nations/people that they intuitively identified as being important in the selection of news. 
The taxonomy of news values has been revised with two main considerations. First, news values were supplemented with a set of commercial news criteria, which were audience-oriented and commercial, by being entertaining and reflecting popular tastes~\cite{golding1979making}. Second, news values were evaluated as to whether they were affected by economic, cultural and political changes in  society. Nearly 40 years after  Galtung's work, Harcup and O'Neil (\citeyear{harcup2001news}) proposed a new set of news values: The Power Elite, Celebrity, Entertainment, Surprise, Bad News, Good News, Magnitude, Relevance, Follow-up, and Newspaper Agenda.

News items that get media attention, however, are dynamic.  In other words, news coverage of an issue rises and falls due to many reasons: change of its news value over time, other news outlets~\cite{dearing1996agenda}, issue competition~\cite{zhu1992issue}, and so on.  
It is important to understand the  dynamics of news coverage correctly because the time span of news items shapes readers' perceptions of the importance of the issue.  
Even high-profile people like
politicians can be influenced when prioritizing issues for decision making~\cite{dearing1996agenda}.

There are mainly two approaches to modeling the temporal dynamics of news coverage.  One is a qualitative modeling. Recently, Boydstun proposed a hybrid model to explain a bi-modal news production process~\cite{boydstun2013making} where the ``alarm'' mode is that news media move quickly from one topic to another, with no in-depth coverage, and the ``patrol'' mode is that news media keep some issues under constant surveillance and produce news on them.  
The other is a quantitative modeling.  Time-series modeling of news coverage on specific issues by tuning parameters is a major approach in this category~\cite{hollanders2008telling}.  

The two closely related topics,  ``what gets media attention'' and ``how the coverage changes over time'',  have been studied for decades. However, those studies are limited by the effort of manual coding; thus, they are conducted with a few example cases or a handful of topics. In this work, using a large-scale news dataset compiled from 196 countries, we answer the following three questions: 1) What gets media attention and is there a global tendency? 2) What is the empirical attention span of news topics across the countries and what gets  long media attention? and 3) Do news topics have different temporal evolution patterns and, if so, then can media attention be characterized by those patterns? 
Our findings will be a valuable asset for the following research to model media attention dynamics.

\section{Data Collection}

Unfiltered News\footnote{http://unfiltered.news/} was developed by Google Ideas (now Jigsaw) to show the most popular topics and the least mentioned topics across the countries. 
Using state-of-the-art machine translation techniques, Unfiltered News successfully indexes all the news articles from the world in many languages. 
For example, Korea, whose official  language is Korean, has its popular topics in English through machine translation.

We collect the most popular 100 topics in 196 countries that are available in Unfiltered News from 7 March to 9 October 2016~\cite{an2017pfi}.  
However, we find that some countries have less than 100 topics on a certain day. 
In the rest of the paper, we use the parameter $k$, which is the minimum number of available topics in each country every day. For example, when $k=10$, the resulting data includes only countries that have the top 10 topics every day from 7 March to 9 October 2016.

\section{What Gets Media Attention}

In examining what gets media attention, we should be careful in determining ``what.'' If it is too specific, the result simply lists the popular events and loses its  generality. By contrast, if it is too broad, the result cannot show any meaningful pattern. 
Keeping the balance, we focus on what \textit{type} of topics gets attention.

Our dataset contains the metadata of each topic, including what type the topic is. 
For example, ``Barack Obama'' is Person type, and ``South Korea'' is Country type.  
All possible types are defined as a hierarchy~\cite{schema}; SportsTeam is a child of SportsOrganization, which is a child of Organization. 
While a topic can have multiple types, we filter out the types whose child type is also assigned to the topic. In other words, we keep the most fine-grained types only.
Furthermore, we augment the types by dividing the Country type into Domestic and Foreign Country types,  and likewise, the City type to Domestic and Foreign City types.

\begin{figure} [hbt!]
  \begin{center}
  \includegraphics[width=70mm]{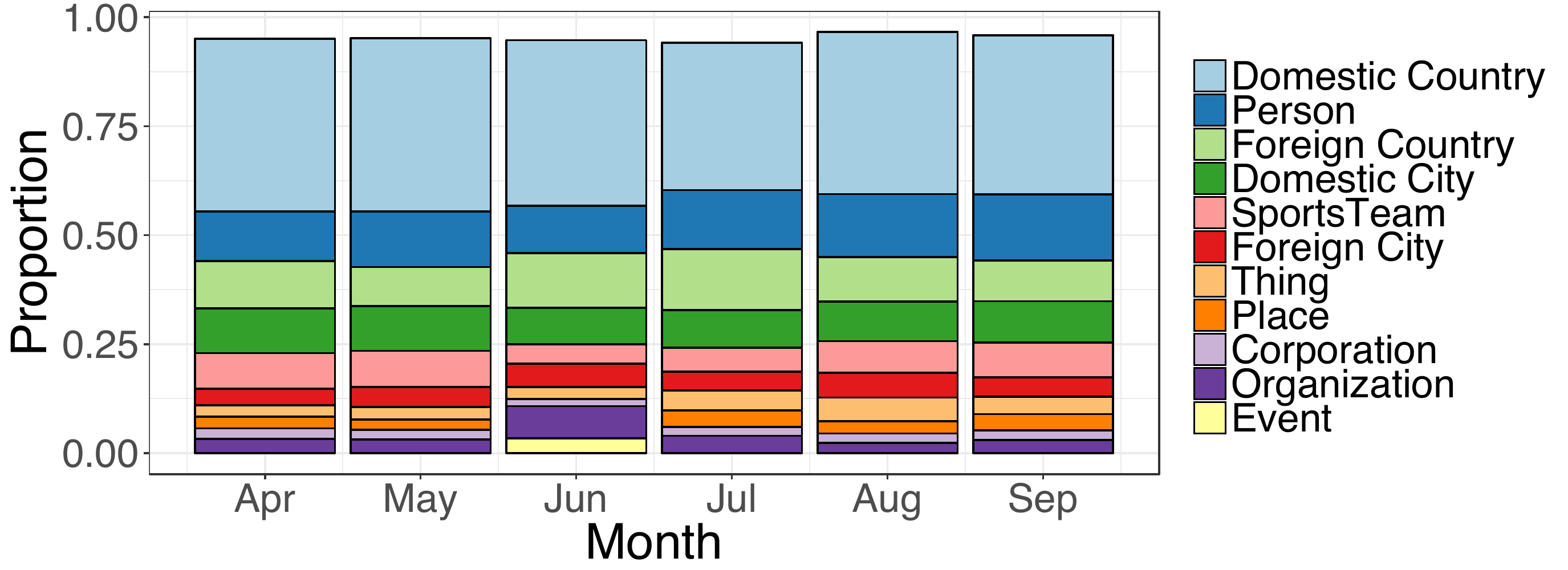}
  \caption{Average profile of topic types}  
  \label{fig:type_proportion}
  \end{center}
\end{figure}

Figure~\ref{fig:type_proportion} shows the average profiles of the top 10 topic types in 129 countries ($k$=10) between April and  September.  Since the number of news articles can be vastly different across the countries, we first compute the proportion of the news articles of each topic type per country, and then calculate their averages across the various  countries.

The average profile of the topic types is quite stable in four out of five months, except for June, when the the Brexit referendum took place (shown as ``Event'' in the figure). The top 10 topic types are consistent, and the average rank correlation coefficient between any two months, except June, is 0.96.  
In other words, unless some global event causes a media storm (a sudden surge of media attention)~\cite{boydstun2014two}, the average order of what gets media attention is relatively stable.

We then examine these topic types one by one.
First, Domestic Country is at the first rank for all five months. As we know anecdotally, what is happening in one's own country is well covered by news media.  The classic news value study explains it by the high relevance of a domestic event to a domestic audience~\cite{harcup2001news}. 
Considering four geographical types, which are Domestic Country, Foreign Country, Domestic City, and Foreign City, the average proportion of domestic ones is 74.1\%.  This is higher than the percentage of domestic news in the '90s, which was 63-66\% measured from 3 media networks in the U.S. for 18 years~\cite{gonzenbach1992world} and thus is well aligned with a recent trend of gradual decrease of foreign news coverage~\cite{altmeppen2010gradual}.

Secondly, Person is constantly ranked second or third in the topic types, which is sensible given that \textit{The Power Elites} and \textit{Celebrity} are two key criteria for news selection~\cite{harcup2001news}.  The power elites indicate not only persons but also organizations and institutions.  In the figure, we see organizations as well.  
Lastly, SportsTeam is also one of the popular topic types. 
The recent taxonomy of news values is increasingly supplemented with factors relating to readers' interests, such as \textit{Celebrity} and \textit{Entertainment}~\cite{harcup2001news}, which explains the coverage of SportsTeam.
As many studies reveal that people enjoy reading sports news~\cite{raney2009handbook}, to attract those readers, on average, 6.8\% of all the news articles published in a country are about sports teams.

\begin{figure} [hbt!]
  \begin{center}
  \includegraphics[width=65mm]{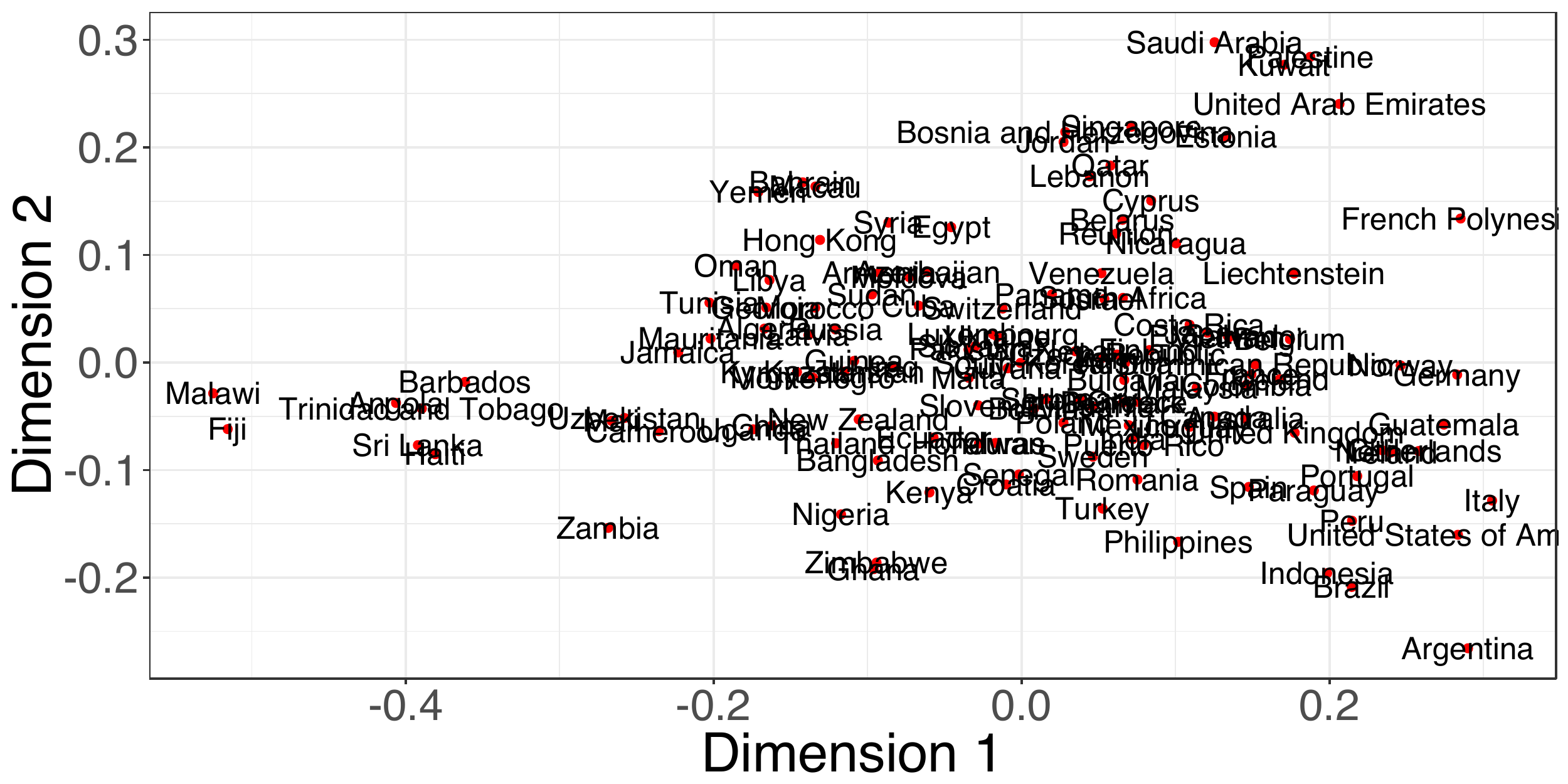}
  \caption{Projection of countries based on their profiles of topic types using classic multidimensional scaling}  
  \vspace{-1mm}
  \label{fig:mds}
  \end{center}
\end{figure}

\begin{figure} [hbt!]
  \begin{center}
  \includegraphics[width=72mm]{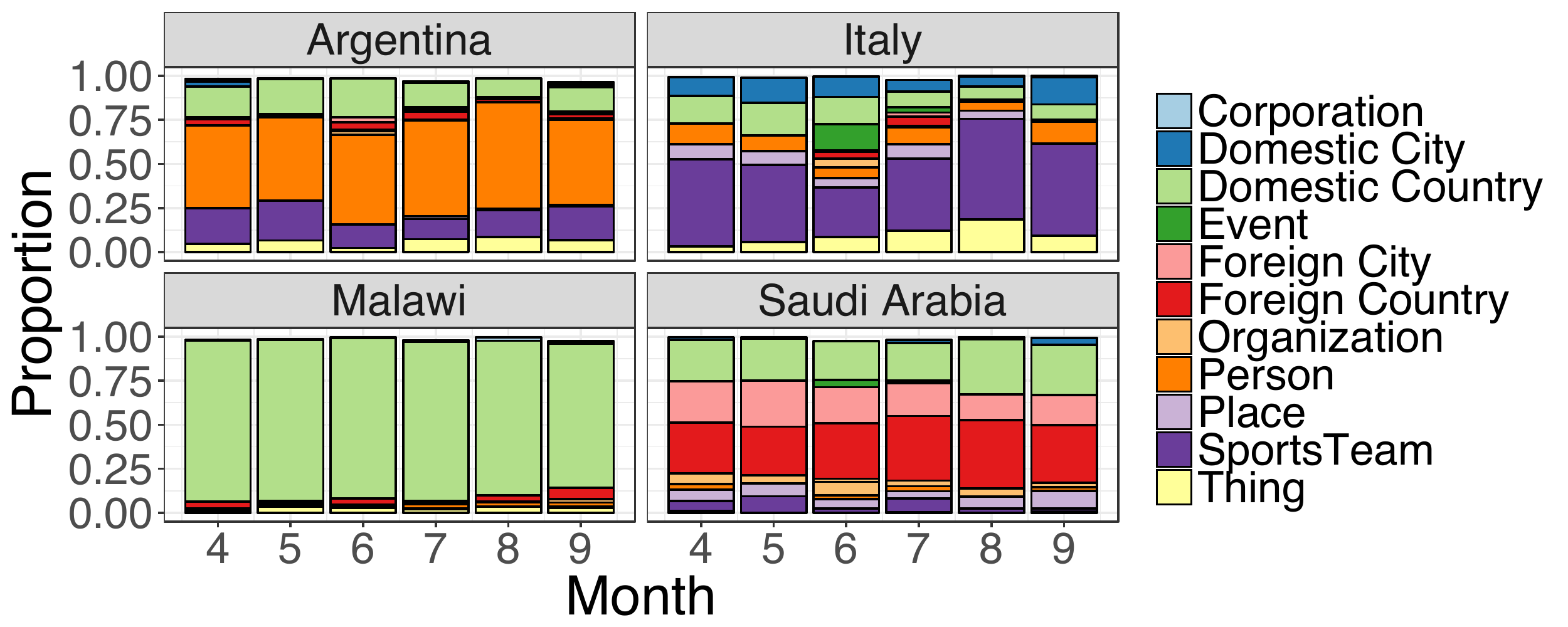}
\vspace{-2mm}    
  \caption{Countries with the most different profiles}  
  \label{fig:type_extreme}
  \end{center}
\end{figure}

While we observed a global profile of topic types as being stable over time; of course, an individual country has a different profile, reflecting its own media attention. 
By using classic multidimensional scaling (MDS), we locate each country in 2-d space in Figure~\ref{fig:mds}. Countries located close to each other are likely to have a similar distribution of topic types in their profiles. To describe what each dimension means, we choose four countries at the farthest points of four directions (left, right, top, and bottom) and Figure~\ref{fig:type_extreme} shows the profiles of those four countries.

The distribution of topic types in the profiles are relatively stable over time within each country, but are markedly different across the countries. For example, at the bottom most position, Argentina has a person-oriented profile;  indicating that as a country is located at the bottom of the plot, its profile becomes similar to that of Argentina.  
The observed differences in the  profiles of topic types demonstrate that what media pay attention to can be notably different across the countries.  
Also, it is interesting that differences in media attention can be captured from the level of topic \textit{type}, not from the level of topic, which is trivial.

In summary, we find a general trend of what gets media attention. Our observations on the geographical relevance, power elites, and sports teams  validate the existing theories on news values by using  worldwide data.  
We also find that the differences in news values across the countries can be observed in the level of topic types.

\section{Temporal Evolution of Media Attention}

In the previous section, we revealed what gets media attention.  Then, how does such attention evolve over time? 
We characterize the evolution of media attention 
by examining: 1) how long media pay attention to a topic (attention span) and 2) the shape of reoccurance attention intensity.

\subsection{Attention Span}

We first measure for how many days in a row a certain topic gets media attention.
We define the attention span toward a topic in country $c$ as the number of consecutive days that the topic is included in the top $k$ topics in the country.

As the average number of unique topics ($k$=10) during 211 days per country is 262.9, simply, the average attention span is 10 (topics per day) $\times$ 211 (days) / 262.9 = 8.0 (days) in an optimistic scenario.  
However, in reality, the attention span in each country is surprisingly low.  When $k$=10, we find that the median of the attention span of all the countries is only 1 day and their average attention span also remains at 2.38 days, 
that is, what media pay attention to, changes quickly.  
Of all the attention spans, the proportion of the attention span of 1 day is 69.3\%.
This indicates that if we see a topic in `the top 10 trending topics' section at a news site today, the probability of seeing the same topic tomorrow is only 1$-$0.693$=$0.307.

Some outliers have longer attention spans, namely, the name of a domestic country or its major cities. As we have seen in an earlier section, a domestic country and city get high media attention. Neighboring countries also tend to have a longer attention span,  because geographical proximity matters in news selection~\cite{sreberny1984world}.

Besides them, we find long attention spans ($>$ 100 days) to do with sports-related topics ($k$=10): Soccer in the Czech Republic (133 days), Germany (131 days), and Slovenia (124 days), and Sport in Egypt (107 days).  
This shows the popular appeal of  sports-related topics.
Also, Terrorism has an attention span of 104 days in Syria, mainly due to ISIS.

\subsection{Temporal Curves of Media Attention}

In the previous section, we observed the short time span of media attention. 
However, as we consider each attention span separately, we lose the information about how media attention recurs.
The recurrent media attention to a topic not only indicates its newsworthiness but also shows general patterns of media attention independent to the current events.
We thus model media attention spans and their recurrence patterns together by generating a time-series for each of the unique 33,911 pairs of $[$country, topic$]$.  The length of each time-series is 211 days, and the magnitude at a certain time point is the number of times that the topic is mentioned in the corresponding country given that day. 

Our aim is to identify representative temporal shapes of the media attention. To this end, in contrast to typical clustering of time-series, it is not important when a peak occurs. Rather, we are more interested in the shape of ``a peak, following the decline, and their recurrence.''
We thus make all the time-series start with a non-zero value by eliminating the first zeros. For this, if we move a time-series by $-$$\Delta$$t_d$, we add $t_d$ zeros to the end of the time-series so that the length of the resulting time-series stays the same for all.
In addition, we do a linear transform of each time-series, mapping to [0,1], to remove the potential bias in the computation of the pair-wise distance of time-series.  We use the dynamic time warping distance to measure the similarity between two time-series. Then, we cluster time-series by K-means++ and choose 4 as the number of clusters by the elbow method.

\begin{figure} [hbt!]
  \begin{center}
  \includegraphics[width=68mm]{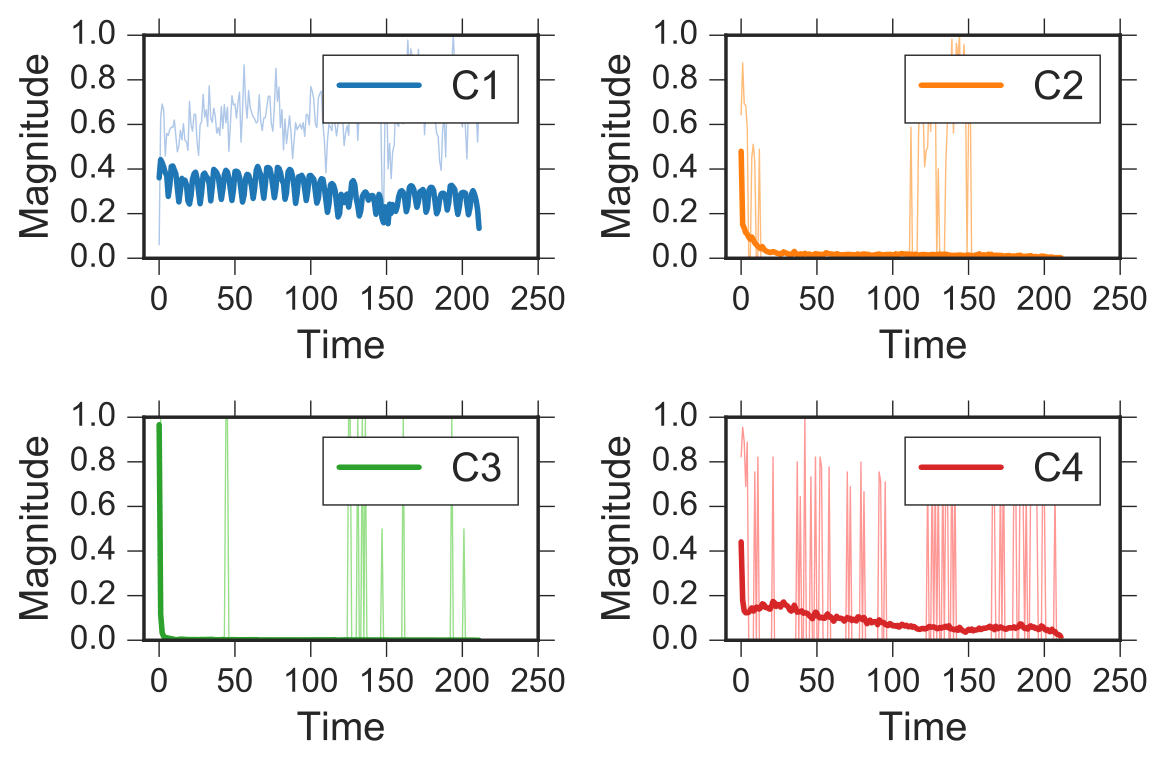}
  \caption{Four clusters of media attention (bold line: centroid, thin line: the real data closest to the centroid)}  
  \label{fig:ts_cluster}
  \end{center}
\end{figure}

Figure~\ref{fig:ts_cluster} shows the four clusters of media attention that have distinct temporal patterns:  Cluster 1 shows a pattern that a topic gets media attention constantly.  Among 33,911 time-series, 752 (2.22\%) show this behavior.  $[$Iraq, Iraq$]$ is a time-series that is closest to the centroid of this cluster.  

A domestic country, which has the highest proportion of topic type in an earlier analysis, falls in this category.  To understand better, we offer examples of the United States of America.  Along with intuitive cases, such as Donald Trump, Hilary Clinton, or NASDAQ, we find that Death and School are also in this category.  For these topics, the media work as watchdogs and produce news continually, functioning as a ``patrol'' system~\cite{boydstun2013making}.   
Cluster 2 is a pattern that whenever a topic gets media attention, it spans multiple days, 
but after that, it rarely gets media attention in contrast to Cluster 1.
16.98\% of the time-series are in this cluster.
In the United States of America, stories on Brexit (e.g. the European Union, United Kingdom European Union membership referendum, etc.) fall in this cluster. Media attention in this cluster is well aligned with the ``alarm/patrol'' mode that produces a burst of news and then continues for a period~\cite{boydstun2013making}.
Cluster 3 is a pattern of characterizing a peak and  fast decline, while a peak might recur. The primary difference between C2 and C3 is whether the peak spans multiple days (C2) or not (C3).  The majority of the time-series (76.32\%) are members of this category.  $[$Barbados, Donald Trump$]$ is closest to the centroid of this cluster.  
Finally, Cluster 4 is a pattern of many sharp peaks. This accounts for 4.48\% of all the time series.  In contrast to C3, the centroid of C4 shows non-zero magnitude over time. Entertainment/Sports news can be in this category because (i) it easily gets much attention from readers,  but (ii) it does not usually lead to the following discussion over days but deals with ephemeral stories (e.g. baseball match results).  $[$Philippines, TV Patrol$]$ is closest to the centroid of this cluster. For topics in C3 and C4, media attention quickly rises and falls, as if it is triggered by an ``alarm''~\cite{boydstun2013making}.

\begin{figure} [hbt!]
  \begin{center}
  \includegraphics[width=68mm]{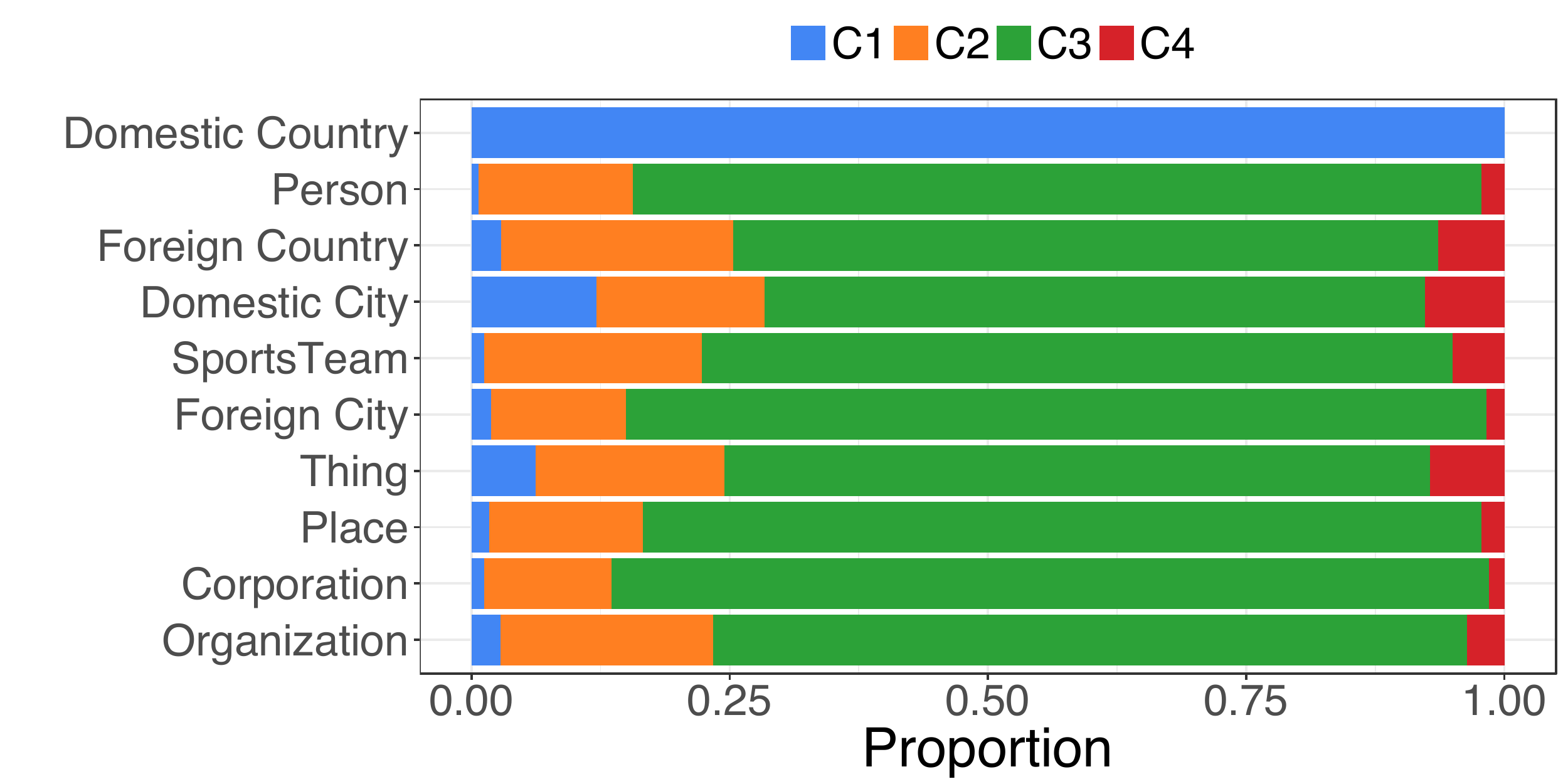} 
  \caption{Proportions of temporal clusters in each topic type}  
  \label{fig:etype_tcluster}
  \end{center}
\end{figure}

Then, how do these temporal clusters correlate with topic types? Figure~\ref{fig:etype_tcluster} shows proportions of the four temporal clusters of each topic type. There are several interesting patterns. First, the temporal evolution of topics of the Domestic Country type is explained by a single temporal pattern, C1, which constantly gets media attention. 
Secondly, 63.9\% (Domestic City) to 84.9\% (Corporation) of topics can be explained by C3, which has a high peak and fast decline.  The short time span of media attention is typical for all the topic types. 
Thirdly, except Domestic Country, Domestic City has the highest proportion of C1 among all other topic types, but its proportion is only 12.1\%.

This indicates that a few domestic cities constantly get media attention, but most of the domestic cities get media attention when it is certainly required.  
Such attention inequality has been mitigated by local media, such as local newspaper or radio, so they focus on local issues. 
Lastly, Foreign Country has the highest proportion (22.4\%) of C2, which has a peak and spans a few days.  The higher proportion of C2 for foreign news is well aligned with the current trend of foreign news coverage, namely that only a few important events are covered,  and thus follow-ups are expected.

In summary, by identifying the four representative temporal shapes of media attention, we discover that news media pay attention to topics differently but have some underlying patterns.  
Most topics have sharp peaks and fast declining attention patterns, reaffirming that the media has a short attention span.

\section{Conclusion}
By analyzing news data from 196 countries, we reveal what gets media attention and how media  attention evolves over time in the modern media system. 
While our findings are well aligned with the previous news value studies, we discover important regional differences in media attention from the perspective of topic types. Also, we propose  four classes of temporal media attention and offer large-scale empirical evidence for the generality of the alarm/patrol media attention model.

\small
\bibliographystyle{aaai}
\bibliography{2017.03-icwsm17-poster-temporal-model}  

\begin{thebibliography}{}

\bibitem[\protect\citeauthoryear{Altmeppen}{2010}]{altmeppen2010gradual}
Altmeppen, K.-D.
\newblock 2010.
\newblock {The gradual disappearance of foreign news on German television: Is
  there a future for global, international, world or foreign news?}
\newblock {\em Journalism Studies} 11(4):567--576.

\bibitem[\protect\citeauthoryear{An and Kwak}{2017}]{an2017pfi}
An, J., and Kwak, H.
\newblock 2017.
\newblock Data-driven approach to measuring the level of press freedom using
  media attention diversity from unfiltered news.
\newblock In {\em the second workshop on news and public opinion (NECO)
  collocated with ICWSM}.

\bibitem[\protect\citeauthoryear{Boydstun, Hardy, and
  Walgrave}{2014}]{boydstun2014two}
Boydstun, A.~E.; Hardy, A.; and Walgrave, S.
\newblock 2014.
\newblock Two faces of media attention: Media storm versus non-storm coverage.
\newblock {\em Political Communication} 31(4):509--531.

\bibitem[\protect\citeauthoryear{Boydstun}{2013}]{boydstun2013making}
Boydstun, A.~E.
\newblock 2013.
\newblock {\em Making the news: Politics, the media, and agenda setting}.
\newblock University of Chicago Press.

\bibitem[\protect\citeauthoryear{Dearing and Rogers}{1996}]{dearing1996agenda}
Dearing, J.~W., and Rogers, E.
\newblock 1996.
\newblock {\em Agenda-setting}, volume~6.
\newblock Sage publications.

\bibitem[\protect\citeauthoryear{Galtung and Ruge}{1965}]{galtung1965structure}
Galtung, J., and Ruge, M.~H.
\newblock 1965.
\newblock The structure of foreign news the presentation of the congo, cuba and
  cyprus crises in four norwegian newspapers.
\newblock {\em Journal of peace research} 2(1):64--90.

\bibitem[\protect\citeauthoryear{Golding and Elliott}{1979}]{golding1979making}
Golding, P., and Elliott, P. R.~C.
\newblock 1979.
\newblock {\em Making the news}.
\newblock Longman Publishing Group.

\bibitem[\protect\citeauthoryear{Gonzenbach, Arant, and
  Stevenson}{1992}]{gonzenbach1992world}
Gonzenbach, W.~J.; Arant, M.~D.; and Stevenson, R.~L.
\newblock 1992.
\newblock The world of us network television news: Eighteen years of
  international and foreign news coverage.
\newblock {\em International Communication Gazette} 50(1):53--72.

\bibitem[\protect\citeauthoryear{Harcup and O'neill}{2001}]{harcup2001news}
Harcup, T., and O'neill, D.
\newblock 2001.
\newblock What is news? galtung and ruge revisited.
\newblock {\em Journalism studies} 2(2):261--280.

\bibitem[\protect\citeauthoryear{Hollanders and
  Vliegenthart}{2008}]{hollanders2008telling}
Hollanders, D., and Vliegenthart, R.
\newblock 2008.
\newblock Telling what yesterday's news might be tomorrow: modeling media
  dynamics.
\newblock {\em Communications} 33(1):47--68.

\bibitem[\protect\citeauthoryear{Ndlela}{2005}]{ndlela2005african}
Ndlela, N.
\newblock 2005.
\newblock The african paradigm: The coverage of the zimbabwean crisis in the
  norwegian media.
\newblock {\em Westminster papers in communication and Culture} 2.

\bibitem[\protect\citeauthoryear{Raney and Bryant}{2009}]{raney2009handbook}
Raney, A.~A., and Bryant, J.
\newblock 2009.
\newblock {\em Handbook of sports and media}.
\newblock Routledge.

\bibitem[\protect\citeauthoryear{Schema.org}{2016}]{schema}
Schema.org.
\newblock 2016.
\newblock \url{https://schema.org}.

\bibitem[\protect\citeauthoryear{Sreberny-Mohammadi}{1984}]{sreberny1984world}
Sreberny-Mohammadi, A.
\newblock 1984.
\newblock The ``world of the news'' study.
\newblock {\em Journal of Communication} 34(1):121--134.

\bibitem[\protect\citeauthoryear{Zhu}{1992}]{zhu1992issue}
Zhu, J.-H.
\newblock 1992.
\newblock Issue competition and attention distraction: A zero-sum theory of
  agenda-setting.
\newblock {\em Journalism \& Mass Communication Quarterly} 69(4):825--836.

\end{thebibliography}

\end{document}